 \documentclass[pmlr,twocolumn,10pt]{jmlr} 

\usepackage{booktabs}
\usepackage{siunitx}
\usepackage[skip=0pt]{caption}
\usepackage{float}

\usepackage[switch]{lineno}

\theorembodyfont{\upshape}
\theoremheaderfont{\scshape}
\theorempostheader{:}
\theoremsep{\newline}

\jmlrvolume{LEAVE UNSET}
\jmlryear{2024}
\jmlrsubmitted{LEAVE UNSET}
\jmlrpublished{LEAVE UNSET}
\jmlrworkshop{ML for Health (ML4H) 2024} 

\title[Random feature baselines]{Random feature baselines provide distributional performance and feature selection benchmarks for clinical and `omic machine learning}

\author{%
\Name{Randall J. Ellis} \Email{randall\_ellis@hms.harvard.edu}\\
\addr Harvard Medical School, United States
\AND
\Name{Audrey Airaud} \Email{audrey.airaud@eleves.enpc.fr}\\
\addr Harvard Medical School, United States
\AND
\Name{Chirag J. Patel} \Email{chirag\_patel@hms.harvard.edu}\\
\addr Harvard Medical School, United States
}

\begin{document}

\maketitle

\begin{abstract}
Identifying predictive features from high-dimensional datasets is a major task in biomedical research. However, it is difficult to determine the robustness of selected features. Here, we investigate the performance of randomly chosen features, what we term ``random feature baselines" (RFBs), in the context of disease risk prediction from blood plasma proteomics data in the UK Biobank. We examine two published case studies predicting diagnosis of (1) dementia and (2) hip fracture. RFBs perform similarly to published proteins of interest (using the same number, randomly chosen). We then measure the performance of RFBs for all 607 disease outcomes in the UK Biobank, with various numbers of randomly chosen features, as well as all proteins in the dataset. 114/607 outcomes showed a higher mean AUROC when choosing 5 random features than using all proteins, and the absolute difference in mean AUC was 0.075. 163 outcomes showed a higher mean AUROC when choosing 1000 random features than using all proteins, and the absolute difference in mean AUC was 0.03. Incorporating RFBs should become part of ML practice when feature selection or target discovery is a goal.

\end{abstract}
\begin{keywords}
genomics; proteomics; supervised ML; feature selection; baselines; benchmarks; UK Biobank
\end{keywords}

\paragraph*{Data and Code Availability}
Code is available on GitHub: \url{https://github.com/RandallJEllis/ml4h_2024/}. 

Data all come from the UK Biobank \citep{sun2023plasma, sudlow2015uk} and cannot be shared due to privacy and security regulations.
\paragraph*{Institutional Review Board (IRB)}
The Harvard internal review board (IRB) deemed the research as non-human subjects research (IRB: IRB16-2145).
\section{Introduction}
\label{sec:intro}

High-dimensional data, including electronic health records, genomics, and imaging, have been used to train ML models for clinical tasks such as diagnosis, prognosis, and biomarker identification. Benchmarking new methods is critical to ML practice. Baseline models are commonly used to compare accuracy between model architectures. Here, we are interested in benchmarking the predictive performance of selected sets of features. 

Human biobanks are an important resource for developing and benchmarking clinical ML methods \citep{all2019all, sun2023plasma, sudlow2015uk}. One avenue of research is to train ML models on blood plasma proteomics data to diagnose and/or predict disease \citep{austin2024plasma, guo2024plasma}. Considerations for developing proteomics-based diagnostic or prognostic assays include predictive performance, cost, and connections to current biological and clinical knowledge. Cost depends on how many proteins are important to quantify to make a correct prediction. For this reason, proteomics-based ML methods often use feature selection to minimize the number of proteins to be measured, while maximizing predictive performance. Ideally, selected proteins will also have prior understanding supporting their use in clinical assays.  

Here, we explore the use of random feature baselines (RFBs) for benchmarking chosen sets of features in clinical ML. We explore two case studies \citep{guo2024plasma, austin2024plasma} using proteomics data from the UK Biobank to predict diagnosis of dementia and hip fracture, respectively. We predict diagnosis of these two conditions using demographics and proteomics alone or in combination, and compare the performance of random sets of proteins of the same size chosen in the original studies (11 and 18, respectively) based on data-driven criteria (Cox regression and Bonferroni correction; SHAP values (\cite{lundberg2020local})). We then examine the performance of RFB when predicting all 607 diseases in the UK Biobank, using 5, 50, 100, 500, or 1000 random proteins, and compare their performance to using all 2,923 proteins in the the dataset.  

\section{Methods}
\subsection{Data preparation}
The UKB Pharma Plasma Proteome (UKB-PPP) cohort consists of blood plasma samples from 54,219 individuals profiled with the Olink Explore 3072 platform, which measures 2,941 protein analytes across 2,923 unique proteins. We removed the patients who were missing basic information such as sex or age. This left 52,956 people, with 1,454 people diagnosed with all-cause dementia (2.7\%), and 692 with hip fracture (1.3\%). To train and test the model, we ran a stratified 80/20 train-test split. For dementia, our demographic variables were age, sex, APOE4 alleles, and education. For hip fracture, they were sex, site, and ethnicity.

\subsection{ML and feature selection}
For dementia and hip fracture RFBs, 100 random sets of 11 and 18 proteins were chosen, respectively. 11 and 18 proteins were selected as these were the same numbers of proteins selected in the original studies \citep{guo2024plasma, austin2024plasma}. When choosing random sets of features, we first removed the proteins chosen in the original studies from the set of proteins to make the results more comparable. We used the autoML framework FLAML \citep{wang2021flaml} for hyperparameter optimization on the training set with a time allotment of 10 minutes, as this exceeded the estimated necessary times outputted by FLAML. We chose our decision thresholds based on maximizing Youden's J-statistic, as done in \citep{guo2024plasma}. We calculated the area under the receiver operating characteristic curve (AUROC), precision, recall, and F1 score for 10,000 bootstraps of the training set and the test set. ML experiments were ran on an internal computing cluster, and no GPUs were used. 

We then used RFB to predict all 607 ``First Occurrence" variables in the UK Biobank, using 5, 50, 100, 500, or 1000 randomly selected proteins, as well as all 2,923 proteins. 100 sets of random features were chosen, each with 100 bootstraps, for a total of 10,000 experiments per number of features per disease. To quantify uncertainty in performance for each set of proteins, we ran 100 bootstraps of the test set, totaling 30,350,000 experiments (5 numbers of proteins x 100 sets x 100 bootstraps x 607 outcomes). We also predicted all 607 outcomes with all proteins, adding an additional 60,700 experiments (100 bootstraps x 607 outcomes). 

\section{Results}
\subsection{Prediction of all-cause dementia}
Table 1 shows the results for predicting all-cause dementia. First, the AUROC of demographics without age was 0.72, and 0.83 including age. This was substantially higher than the mean AUROC of random sets of 11 proteins (0.6), but the combination of both yielded a mean of 0.81, about the same as the chosen 11-protein panel alone in \citep{guo2024plasma}. 

\begin{table*}[htbp]
\floatconts
  {tab:operatornames}%
  {\caption{Predictive performance for all-cause dementia}}%
  {%
  \scriptsize
    \begin{tabular}{lllll}
    \toprule
    Features & AUROC & Precision & Recall & F1 \\
    \midrule
    Demo, No Age & 0.72 (0.71-0.73) & 0.06 (0.05-0.07) & 0.55 (0.51-0.59) & 0.11 (0.1-0.12) \\
    Demographics & 0.83 (0.82-0.84) & 0.08 (0.07-0.09) & 0.72 (0.69-0.75) & 0.15 (0.13-0.16) \\
    Proteins & 0.6 (0.53-0.67) & 0.04 (0.03-0.06) & 0.48 (0.12-0.67) & 0.07 (0.04-0.09) \\
    Proteins+Demo & 0.81 (0.8-0.82) & 0.07 (0.06-0.09) & 0.74 (0.63-0.79) & 0.13 (0.12-0.16) \\
    Guo (P) & 0.84 (0.82-0.86) & N/A & N/A & N/A \\
    Guo (P+D) & 0.91 (0.89-0.93) & N/A & N/A & N/A \\
    \end{tabular}
    \par
    
    \vspace{0.1em}
    \footnotesize{AUROC: Area Under the Receiver Operating Characteristic curve. Demo: Demographics. Guo (P) shows the performance of a chosen 11-protein panel from \citep{guo2024plasma}. Guo (P+D) shows the performance of the same protein panel and demographics (age, sex, education, APOE4 alleles).}
  }
\end{table*}
Unfortunately, only AUROC was reported in the original study, so we cannot make comparisons on the other metrics.

\subsection{Prediction of hip fracture}
Table 2 shows the results for predicting hip fracture. First, the AUROC of demographics was 0.6. This was lower than the mean AUROC of random sets of 18 proteins (0.67), with substantial gains in recall. The combination of both yielded a mean of 0.64, which was 0.125 less than the performance of the best model from \citep{austin2024plasma}. 

\begin{table*}[htbp]
\floatconts
  {tab:operatornames}%
  {\caption{Predictive performance for hip fracture}}%
  {%
  \scriptsize
    \begin{tabular}{lllll}
    \toprule
    Features & AUROC & Precision & Recall & F1 \\
    \midrule
    Demographics & 0.6 (0.58-0.61) & 0.04 (0.01-0.07) & 0.04 (0.01-0.07) & 0.04 (0.01-0.07) \\
    Proteins & 0.67 (0.6-0.75) & 0.02 (0.01-0.03) & 0.59 (0.29-0.89) & 0.04 (0.02-0.06) \\
    Proteins+Demo & 0.64 (0.58-0.70) & 0.02 (0.01-0.03) & 0.5 (0.21-0.78) & 0.04 (0.03-0.06) \\
    Austin & 0.765 & N/A & N/A & N/A \\
    \end{tabular}
    \par
    
    \vspace{0.5em}
    \footnotesize{AUROC: Area Under the Receiver Operating Characteristic curve. Demo: Demographics. Austin shows the performance of a combination of a clinically derived risk score (FRAX-CRF) and a proteomics derived risk score from \citep{austin2024plasma}.}
  }
\end{table*}

\begin{figure}[htb]
    \centering
    \includegraphics[width=0.4\textwidth, trim={0pt 80pt 570pt 0pt}, clip]{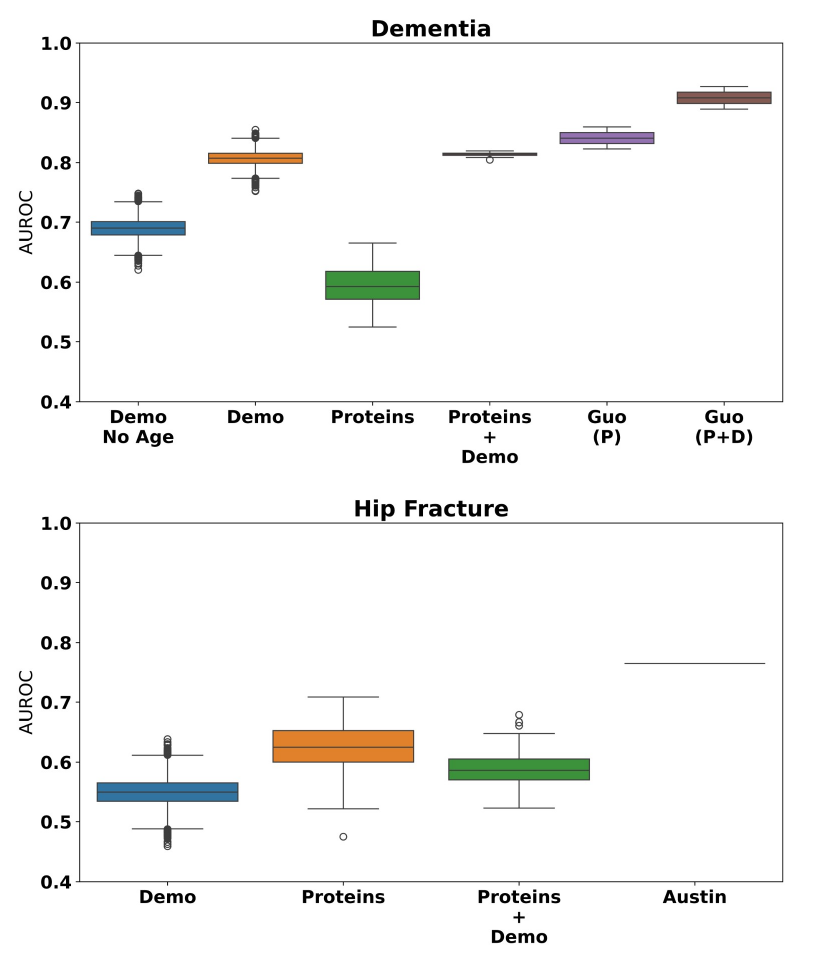}
    \caption{Box plots showing the area under the receiver operating characteristic curve for predicting dementia and hip fracture, for demographics, RFBs, their combination, as well as the performance from the original publications. For Dementia - Demo No Age: Demographics, but without age (i.e., sex, APOE4 alleles, education); Demo: Age, sex, APOE4 alleles, education; Proteins: 100 randomly chosen sets of 11 proteins (the same number chosen in \cite{guo2024plasma}), each with 100 bootstraps; Proteins+Demo: Same 100 sets of randomly chosen proteins with the addition of demographics; Guo (P): 11-protein panel from \cite{guo2024plasma}; Guo (P+D): 11-protein panel with demographics. For hip fracture - Demo: sex, assessment site, ethnicity; Proteins: 100 randomly chosen sets of 18 proteins (the same number chosen in \cite{austin2024plasma}), each with 100 bootstraps; Proteins+Demo: Same 100 sets of randomly chosen proteins with the addition of demographics; Austin; Performance of a combination of a clinically derived risk score (FRAX-CRF) and a proteomics derived risk score from \cite{austin2024plasma}.}.
    \label{fig_boxplots}
    
\end{figure}

Again, only AUROC was reported in the original study, so we cannot make comparisons for other measures. It is not clear how decision thresholds were chosen in the original study, so we cannot make assumptions about the precision, recall, and F1 of the model developed in the original study. \autoref{fig_boxplots} shows the AUROC results graphically, by model configuration.

\subsection{Prediction of  607 ``First Occurrence" variables in the UK Biobank}
We then used RFB to predict all 607 ``First Occurrence" disease variables in the UK Biobank. We randomly chose 5, 50, 100, 500, or 1000 proteins (100 random sets of each number) to predict each disease, along with all 2,923 proteins in the dataset.

\begin{figure}[htb]
    \centering
    \includegraphics[width=0.5\textwidth, trim={0pt 100pt 0pt 70pt}, clip]{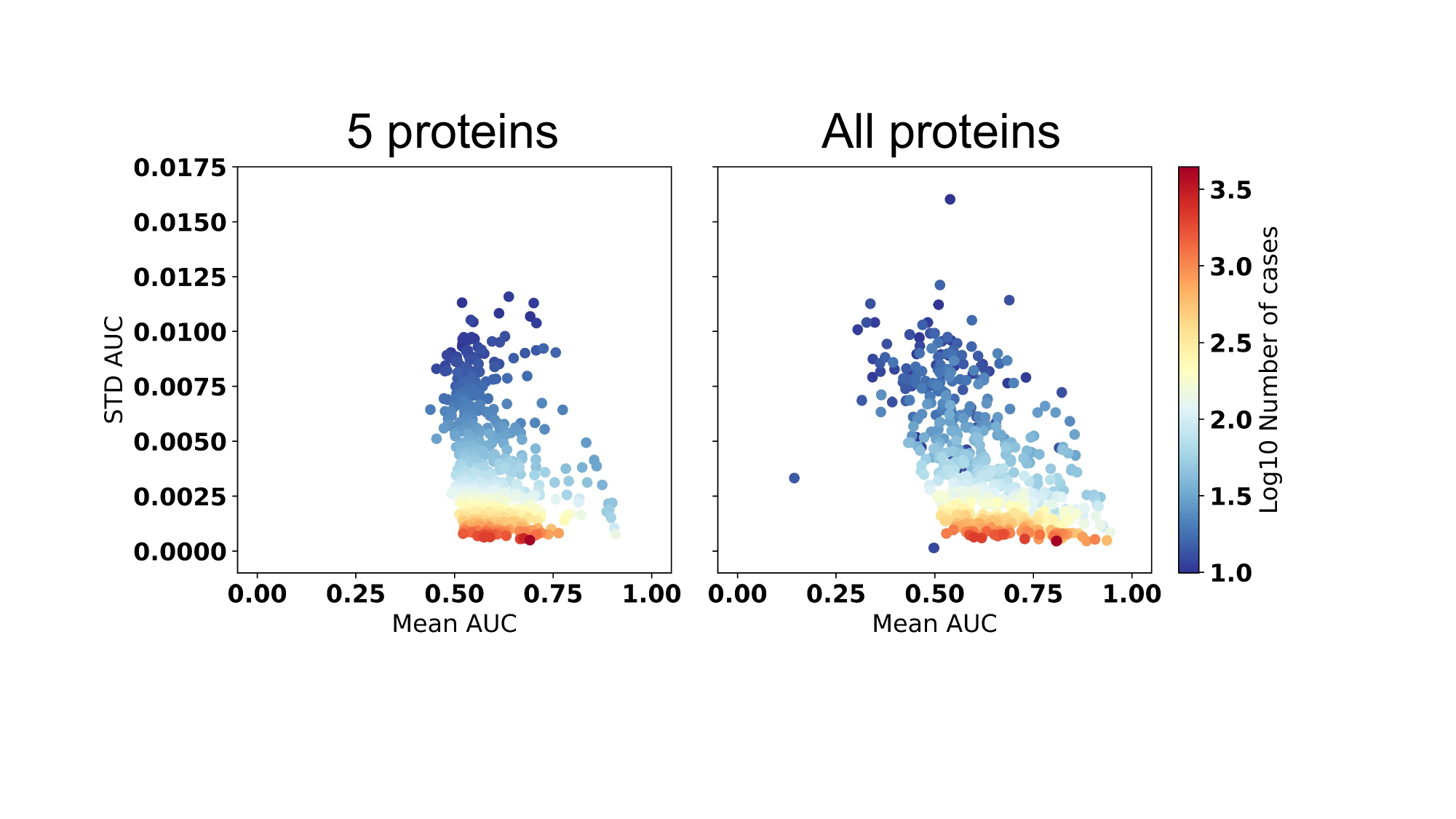}
    \caption{Mean and standard deviation of the area under the receiver operating characteristic curve for predictions of 607 disease outcomes in the UK Biobank, using 100 sets of 5 randomly chosen proteins, or all 2,923 proteins in the proteomics dataset.}
    \label{fig_5_vs_all}
\end{figure}

For most outcomes, the difference in mean AUROC was marginally different between 5 randomly selected proteins and all proteins, with variance in performance being anticorrelated to the sample size of cases (\autoref{fig_5_vs_all}).

\begin{figure}[htb]
    \centering
    \includegraphics[width=0.6\textwidth, trim={70pt 20pt 0pt 0pt}, clip]{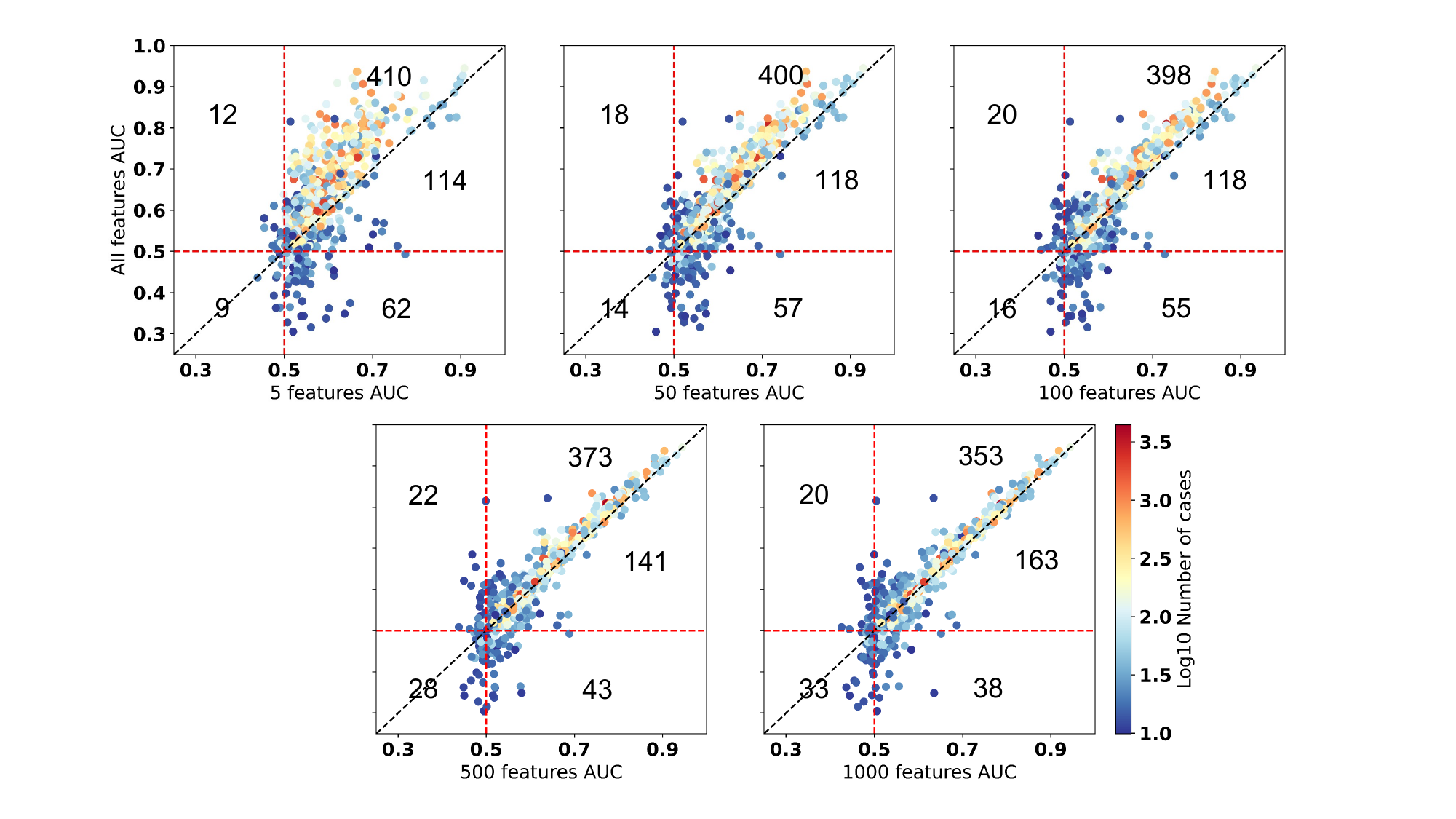}
    \caption{Comparisons of mean AUROC between 5, 50, 100, 500, 1000, and all proteins for predicting 607 disease outcomes in the UK Biobank. The red dashed lines denote 0.5 (i.e., chance performance), and black dashed lines denote equal performance. The numbers denote, starting from the top left: outcomes with a mean AUROC\textgreater 0.5 when using all proteins, but \textless 0.5 for the RFB; outcomes where both conditions have AUROC\textgreater 0.5, but is higher when using all proteins; outcomes where both conditions have AUROC\textgreater 0.5, but is higher when using the RFB; outcomes with a mean AUROC\textgreater 0.5 with the RFB, but \textless 0.5 using all proteins; outcomes where AUROC\textless 0.5 for both conditions}.
    \label{5to1000_vs_all}
\end{figure}

We compared the mean AUROC between each number of randomly chosen proteins and all proteins, for all 607 outcomes (\autoref{5to1000_vs_all}). 

One surprising finding was that 114/607 outcomes showed a higher mean AUROC when choosing 5 random features than using all 2,923 proteins, and the mean absolute difference in AUC across outcomes was 0.0749. The three outcomes with the greatest mean AUC difference, favoring 5 random proteins were antepartum haemorrhage, not elsewhere classified (mean difference: 0.244), secondary parkinsonism (0.182), and other complications of labour and delivery, not elsewhere classified (0.173). The three outcomes most favoring all proteins are aspergillosis (0.301), insulin-dependent diabetes mellitus (0.288), and intestinal malabsorption (0.274). 163/607 outcomes showed a higher mean AUROC when choosing 1000 random features than using all proteins, and the mean absolute difference in AUC was 0.0292. The three outcomes with the greatest mean AUC difference, favoring 1000 random proteins were gestational [pregnancy-induced] hypertension with significant proteinuria (mean difference: 0.174), obstructed labour due to malposition and malpresentation of foetus (0.137), and preterm delivery (0.133). The three outcomes most favoring all proteins are aspergillosis (0.311), insulin-dependent diabetes mellitus (0.187), and intestinal malabsorption (0.135).
\vspace{5pt}

\vspace{-20pt}

\section{Discussion}
Developing and optimizing predictive models as combinations of algorithms, features, and other choices, requires testing baselines to demonstrate the value of these choices. While it is common to test different algorithms, it is uncommon to use randomly chosen features as baselines. Randomly chosen features are a valuable baseline in contexts when specific features are selected for a particular purpose and where there is dense correlation between features. In clinical ML, where features are selected based on biological and clinical knowledge, as well as predictive performance, it is critical to compare selected features to random features to ensure that valuable signal is detected by the selected set of features. Additionally, when features are selected from an 'omic or other high-dimensional data source and combined with more common or standard modalities such as demographics or causal variables (the clinical state of the art), it is essential to show the performance of demographics alone to demonstrate the performance gained by adding features from the 'omic modality. 

In the two case studies we examined, performance achieved with demographics alone, or demographics with randomly chosen proteins was comparable to that  of the chosen and optimized models and feature sets reported in the papers. Both studies only reported AUROC, and other measures are worth reporting, particularly when a class imbalance is present. Assuming the precision, recall, and F1 achieved by our models is comparable to that achieved by the models developed in the original studies, we point out that precision may be particularly low. Therefore,   a majority of participants predicted by these models to have a disease do not. This is a significant point for assessing the clinical value of a predictive model, and one that is sometimes obfuscated by only reporting AUROC, and moreso in the presence of a class imbalance. 

For ML workflows including feature selection, RFBs are an important benchmark to confirm the value of selected features. For biomarker development, RFBs are a test of the predictive performance of the candidate biomarker.  

A limitation of this work is the non-exhaustive testing of different numbers of randomly chosen features. We also will investigate in future work the correlation structure of the proteins, as this likely plays a role in the large portion of performance captured by randomly chosen features.

\newpage
\bibliography{jmlr-sample}

\end{document}